\documentclass{iopart}
\usepackage{graphicx}
\usepackage{amstext}
\usepackage{amsfonts}
\usepackage{amssymb}
\newcommand{\be}{\begin{equation}}
\newcommand{\ee}{\end{equation}}
\newcommand{\M}{\mathcal{M}}
\newcommand{\PF}{\mathcal{P}}

\begin{document}

\review{Resonances in open quantum maps}

\author{Marcel Novaes}
\address{Departamento de F\'isica, Universidade Federal de
S\~ao Carlos,\\ S\~ao Carlos, SP, 13565-905, Brazil}

\begin{abstract}

We review recent studies about the resonance spectrum of quantum scattering systems, in
the semiclassical limit and assuming chaotic classical dynamics. Stationary quantum
properties are related to fractal structures in the classical phase space. We focus
attention on a particular class of problems that are chaotic maps in the torus with
holes. Among the topics considered are the fractal Weyl law, the formation of a spectral
gap and the morphology of eigenstates. We also discuss the situation when the holes are
only partially transparent and the use of random matrices for a statistical description.

\end{abstract}



\section{Introduction}

A closed quantum system has a discrete spectrum of energy levels, to which are associated
bound states. Open systems, on the other hand, have a resonance spectrum, consisting of a
discrete set of complex poles of the Green's function or of the scattering matrix
(besides having a real continuous spectrum of scattering states). If we imagine a closed
system being slowly opened, its bound states will acquire small (negative) imaginary
parts and become resonances. In this sense, the resonance spectrum can also be seen as
eigenvalues of a non-hermitian Hamiltonian. If $\mathcal{E}=E-i\Gamma/2$ is a complex
energy value, the usual time evolution factor $|e^{-i\mathcal{E}t/\hbar}|^2=e^{-\Gamma
t/\hbar}$ is not unimodular but decreases exponentially. The quantity $\Gamma/\hbar$ is
thus called the decay rate of the state, or equivalently $\hbar/\Gamma$ is called its
lifetime. Insofar as any real system is always in contact with its environment, most
energy levels are actually resonances with a finite decay rate.

We are interested in systems for which the classical dynamics is chaotic. Scattering of
waves in chaotic systems has been experimentally realized in a great variety of physical
systems. Out of the vast literature available, we select only a few examples: electron
transport in semiconductor quantum dots \cite{dots1,dots2,dots3,dots4} and graphene
quantum dots \cite{graphene1,graphene2,graphene3}; microwaves in normal metallic
\cite{microwaves1,microwaves2,microwaves3} or superconducting cavities
\cite{microwaves4}; nuclear reactions \cite{nuclear1,nuclear2}; acoustic waves
\cite{acoustic}; microlasers \cite{lasers1,lasers2,lasers3}.

In the semiclassical limit $\hbar\to 0$ the wavelength is much smaller than any classical
length scale and the nature of the classical (or ray) dynamics becomes important. Three
main questions then stand out in connection with the resonance spectrum of a chaotic
scattering system. First, how does the number of resonances with given decay rate grow
with $\hbar$? Second, is there a lower bound for the decay rates? Third, what do the
resonance wavefunctions look like (inside the scattering region)? As we will see, all
these questions are related to fractal properties of the classical dynamics. They are all
still open to some extent, although much progress has indeed been made.

The systems we have in mind have `classical' openings. By this we mean the following. For
a fixed value of the energy, a finite `energy shell' can be constructed by imagining a
big enough box enclosing the scattering region in configuration space and all possible
momenta with fixed magnitude. In this energy shell, the points leading to immediate
escape occupy a finite volume. In this situation the resonances are always strongly
overlapping (sometimes called the Ericson regime).

A very popular class of models with which one can theoretically study chaotic scattering
are so called torus maps, some of which are well known like the baker map or cat maps.
These toy models have the maximal simplicity still allowing for chaotic dynamics. Also,
they can be quantized in order to study wave properties. In the last few years, the
resonance spectrum of open quantum maps has been widely studied and have shed
considerable light on the more general problem. Our purpose is to review these
developments.

Other reviews concerning areas related to the ones discussed here have appeared recently
and can offer complementary insights. Open systems and non-hermitian Hamiltonians were
reviewed by Rotter in \cite{Rotter} and, from a very different point of view, by de la
Madrid and Gadella in \cite{madrid} (see also \cite{bohm}); open billiards were reviewed
by Dettmann in \cite{Dettmann} and also by Altmann, Portela and T\'el in \cite{leaking};
the random matrix theory (RMT) approach to resonances in quantum systems was reviewed by
Fyodorov and Savin in \cite{SavinFyod} and by Mitchell, Richter and Weidenm\"uller in
\cite{RevModPhys}. Finally, spectral results for chaotic open maps were reviewed by
Nonnenmacher in \cite{NonRev}, but from a mathematically more rigorous perspective than
the one adopted here.

In Section 2 we introduce open chaotic maps. We give a few examples and discuss their
long time behaviour. We introduce fractal trapped sets and conditionally invariant
measures. In Section 3 we discuss open quantum maps and their general properties. Section
4 is dedicated to the fractal Weyl law that has been conjectured to hold for the
resonance spectrum in chaotic scattering. Section 5 discusses the possible existence of a
spectral gap. The phase space morphology of resonance wavefunctions is considered in
Section 6. In Section 7 we focus on systems where escape is not ballistic, like e.g.
microlasers. The RMT approach to quantum scattering is discussed in Section 8. Finally,
we close in Section 9 with some open problems.

Let us mention that we do not consider the closely related, and somewhat complementary,
transport formulation of scattering \cite{transport}, which deals with $S$ matrices,
transmission eigenvalues, counting statistics, etc. The semiclassical and the RMT
approaches to quantum chaotic transport were reviewed, for instance, in
\cite{semiclassical} and \cite{beenakker}, respectively.

\section{Classical maps}

When studying the integrability of the dynamics of the solar system, Poincar\'e
introduced the idea of `cutting through' phase space with a hyperplane, and recording the
intersections of the system with this plane. This construction, now known as a Poincar\'e
section, produces a discrete-time dynamics in which the trajectory becomes a sequence of
points. The time interval between intersections is lost, but this is usually considered a
small price to pay for the dimensional reduction achieved.

We consider maps in themselves, without any reference to a continuous-time system from
which they could be derived. We shall treat only maps that are defined on a
two-dimensional torus, which can be represented simply as a square with opposite sides
identified. Let us denote by $(q,p)$ orthogonal coordinates in this space which are the
sides of the square. We have in mind conservative chaotic dynamics (hyperbolic and
mixing).

Let $\M$ denote the map. If $(q',p')=\M(q,p)$, we say that the point $(q,p)$ evolves into
the point $(q',p')$. An infinite sequence of points is called an orbit. If an orbit
consists in the infinite repetition of a finite number of distinct points, it is called
periodic. The number of distinct points is then its period. Its is known that chaotic
systems have infinitely many periodic orbits, and that they form a dense set in phase
space, i.e. there is a periodic orbit arbitrarily close to any point.

Many books discuss chaotic maps. We just mention for example \cite{Dorfman,Ott}, where
all basic definitions, examples and further discussion can be found. In the following, we
briefly present a few paradigmatic systems and introduce only the notions that will be
needed when we consider the resonances of open quantum maps.

Perhaps the simplest example is the baker map, given by \be (q ,p)\mapsto\cases{ (2q,
p/2),& if $q<1/2$,\cr (2q-1, (p+1)/2), & if $q>1/2$.}\ee Its well-known dynamics consists
in a stretching by a factor of $2$ along the $q$ axis, together with a contraction by the
same amount in the $p$ axis, plus a `cut and paste' operation involving the line $q=1/2$.
Clearly, area is preserved. Also, the map is hyperbolic, with stable and unstable
directions at every point being parallel to the coordinate axes. The Lyapounov exponent
is $\ln 2$. Notice that the baker map may be written as \be (q ,p)\mapsto (2q-[2q],
(p+[2q])/2),\ee where $[2q]$ denotes the integer part of $2q$.

Many generalizations of the baker map exist. One of them makes use of $D$ basic regions,
stretching and compressing by a factor of $D$:
 \be (q,p)\mapsto (Dq-[Dq], (p+[Dq])/D).\ee This is sometimes called the
$D$-nary baker map. The usual baker map corresponds to $D=2$. This map has a Markov
partition with $D$ cells, and each cell is stretched and compressed by a factor $D$ at
each iteration. Therefore, the Lyapounov exponent in this case is $\ln D$.

The so-called cat maps provide another paradigmatic family of chaotic linear maps. The
following is an example: \be\label{cat} (q ,p)\mapsto (2q+p, 3q+2p) {\rm mod} 1.\ee In
this case the Lyapounov exponent is $\ln(2+\sqrt{3})$. The dynamics of cat maps has no
discontinuity. The stable and unstable directions are still the same at every point, but
they are no longer orthogonal.

Our final example will be the standard map. It is defined as \be (q,p)\mapsto
(q+p+K\sin(2\pi q),p+K\sin(2\pi q)).\ee It is more generic than the previous examples
because it is non-linear. Its dynamics is known to be predominantly chaotic if $K>7$. The
stable and unstable directions, as well as the local stretching factor, are no longer
constant in phase space. It is known that the Lyapounov exponent (the average stretching
factor) is given approximately by $\ln(K/2)$.

Instead of propagating points, it is often useful to consider propagation of probability
measures. Given a probability measure $\mu$, its evolution is another probability measure
$\mu'$ such that, for any subset $A$ of phase space, $\mu'(\M(A))=\mu(A)$. A measure
$\mu$ is called {\em invariant} if $\mu(\M(A))=\mu(A)$ for every $A$. If we allow
singular invariant measures, such as linear combinations of Dirac deltas, then clearly
there are infinitely many of them (for example, to any periodic orbit we can associate
one).

If a probability measure $\mu$ has a probability density $f$, such that $\mu(A)=\int_A
f(q,p)dqdp$, it is called absolutely continuous. Probability densities can also be
evolved in time: if $(q',p')=\M(q,p)$, then the function $g(q,p)$ such that
$g(q',p')=f(q,p)$ is called the evolution of $f$ (remember that the map is conservative,
so there is no Jacobian). In other words, each point simply carries along the value of
the function associated with it. The linear operator $\PF$ that implements this
evolution, $g=\PF f$, is called the Perron-Frobenius operator associated with the map
$\M$.

All examples of chaotic maps discussed previously are ergodic systems and hence have only
one invariant probability density. This is called the natural density or equilibrium
density, denoted $f_e$ (for conservative maps this is simply the constant function). They
also have the exponential mixing property, which implies that any smooth initial density
function will converge to $f_e$ exponentially fast in time (when defined on an
appropriate function space, $\PF$ has a single eigenvalue equal to $1$ and all other
eigenvalues have strictly smaller modulus). This convergence of course takes place in a
weak sense.

We now wish to consider open maps as models of chaotic scattering. Given a map, an open
version is obtained by defining a `hole' in phase space, which can in principle be any
region of finite area (or union of regions). Once a point falls into the hole, it stops
being propagated (i.e. it `escapes'). This means that the dynamics is no longer
conservative. In fact, for chaotic systems almost all initial conditions eventually
escape because of ergodicity. We denote by $\widetilde{\mathcal{M}}$ the open map, i.e.
the map that acts by first removing points in the hole and then evolving the remaining
ones according to $\mathcal{M}$.

The trapped set of the dynamics is the set of initial conditions which do not originate
from the hole and never hit the hole. In other words, they remain in the system for
infinite time, whether propagated forwards or backwards. This invariant set will be
denoted $K_0$. It is quite common in the quantum chaos literature to refer to $K_0$ as
the system's `repeller'. This has been criticized on the grounds that `repeller' should
be used for sets that are unstable in all directions. When a stable manifold exists,
\cite{leaking} and \cite{telbook} suggest calling $K_0$ the {\em chaotic saddle} or
simply the `saddle'. We shall follow this suggestion.

The stable manifold of the saddle, the set of points which converge to it and therefore
never escape in the future, is also known as the forward-trapped set and denoted $K_+$.
On the other hand, the unstable manifold contains the initial conditions which never
escape in the past and is also called the backward-trapped set $K_-$. These two sets are
fractals and have similar structures. Globally, they are very convoluted. Locally, they
are continuous in one direction and fractal in the orthogonal one. Therefore, they both
have zero measure in phase space.

The sets $K_-$ and $K_+$ can be described in terms of images and pre-images of the hole.
Let $h^m=\mathcal{M}^m h$ be the $m$th image of the hole under the dynamics. The set
$h^0=h$ is the hole itself. These sets can have finite intersections, so we define
another family of sets as $\mathcal{R}^m=\widetilde{\mathcal{M}}^m h$. The set
$\mathcal{R}^m$ contains the points that fall into the hole in exactly $m$ steps when
propagated backwards. Analogously, we can start with $\mathcal{R}^{0}=h$ and for $m>0$
define $\mathcal{R}^{-m}$ as the set of points which fall into the hole in exactly $m$
steps when propagated forwards. These sets obey
$\mathcal{M}\mathcal{R}^{m}\supset\mathcal{R}^{m+1}$, and \be\label{Rs}
\widetilde{\mathcal{M}}\mathcal{R}^{m}=\mathcal{R}^{m+1}.\ee

\begin{figure}[t]
\includegraphics[]{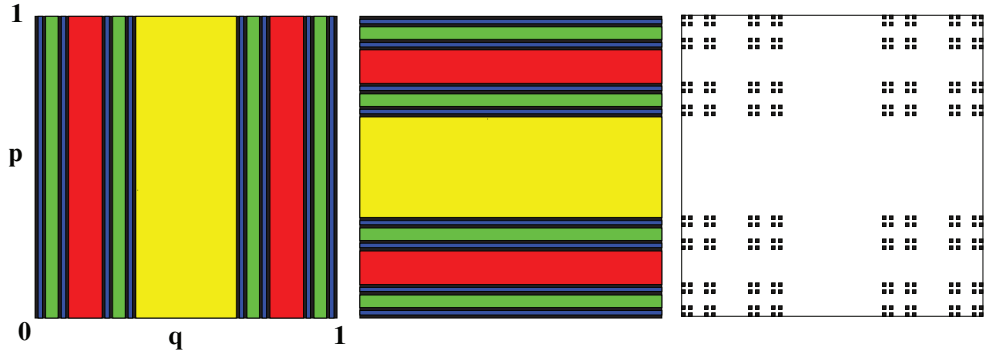}
\caption{(color online) Important sets for the open ternary baker map. The region
$1/3<q<2/3$ corresponds to the hole. The set of points that fall in the hole in exactly
one step ($\mathcal{R}^{-1}$) have $q$ in the region $(1/9,2/9)\bigcup (7/9,8/9)$. The
darkest lines in the left panel are an approximation to the forward trapped set, $K_+$.
In the middle panel we see the sets $\mathcal{R}^{1}$ as $1/3<p<2/3$, $\mathcal{R}^{2}$
as $p\in(1/9,2/9)\bigcup (7/9,8/9)$,etc. The darkest lines in this panel approximate
$K_-$. The right panel shows a finite-time approximation to the invariant saddle $K_0$,
which is the product of two Cantor sets. Figure reprinted with permission from
\cite{NZ1}.}
\end{figure}

\begin{figure}[t]
\includegraphics[scale=2]{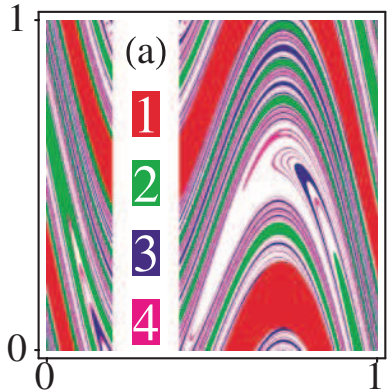}
\caption{(color online) Similar to the left panel of Figure 1, but for the kicked
rotator. Different colors (shades) correspond to points that escape in one, two, three
and four steps. The white region is a finite-time approximation to $K_+$. Figure
reprinted with permission from \cite{henning}. Copyright (2004) by the American Physical
Society.}
\end{figure}

Let us imagine that we start with the torus completely filled with initial conditions,
and they are all propagated forwards. First, the points at the hole escape. Then, the
points that were initially at $\mathcal{R}^{-1}$ escape. Next, the points initially at
$\mathcal{R}^{-2}$, and so on. Clearly, since it contains the points that never escape,
the forward trapped-set $K_+$ must be the complement of the union of all
$\mathcal{R}^{-m}$. On the other hand, the backward-trapped set $K_-$ is the complement
of the union of all $\mathcal{R}^{m}$. Notice that the sets $\mathcal{R}^{-m}$ intersect
the unstable manifold $K_-$ and in fact provide a useful partition of it (every point of
$K_-$ is either in one of the $\mathcal{R}^{-m}$ or belongs to $K_0$).

The simplest example of open map with a nontrivial saddle is the ternary baker map with
the middle region identified with the hole. This is discussed in detail, for example, in
\cite{NZ1}. It is easy to see that all sets $\mathcal{R}^{-m}$ consist of vertical
strips. The horizontal section of $\mathcal{R}^{-1}$, for example, is
$(1/9,2/9)\cup(7/9,8/9)$. The set $\mathcal{R}^{-m}$ is always the union of $2^m$ strips
of width $3^{-m}$. As a matter of fact, the forward trapped set is the product of the
interval $[0,1]$ in the vertical direction and a Cantor set in the horizontal direction.
Analogously, the backward trapped set is a Cantor set in the vertical direction. Finally,
the saddle $K_0$ is the product of two Cantor sets. This example can be seen in Figure 1,
while the analogous construction for the kicked rotator is shown in Figure 2.

Open systems can have invariant measures, but they must be supported on the saddle and
thus are very singular. Let $\mu^t$ denote the $t$-step propagation of measure $\mu$. A
measure satisfying $\mu^t=e^{-\Gamma t}\mu$ is called an eigenmeasure of the system.
Invariant measures correspond to $\Gamma=0$.

Incidentally, open systems provide a glimpse at how truly vast is the set of invariant
measures for closed ones. A closed system can be opened by choosing any region of phase
space as the hole. In turn, this may generate a chaotic saddle, which can carry many
invariant (singular) measures. These are also invariant measures of the closed system,
and we thus have many such measures for many possible choices of hole in phase space.

Eigenmeasures with $0<\Gamma<\infty$ are called conditionally invariant measures (in the
sense that they are invariant provided we renormalize them at every step).  Conditionally
invariant measures for maps were reviewed in \cite{young}. They are necessarily supported
on $K_-$. Since this set has a fractal `cross section', they cannot be absolutely
continuous. If we take smooth initial measures and propagate them forwards, they all
still converge (when renormalized) to the same measure $\mu_e$, called the natural
measure or equilibrium measure. It is an eigenmeasure satisfying
\be\label{equilibrium}\mu_e^t=e^{-\gamma t}\mu_e.\ee

The quantity $\gamma$ appearing in (\ref{equilibrium}) is called the system's decay rate.
Decay rates for chaotic systems with holes have been reviewed in \cite{leaking}. In the
limit of vanishingly small holes, $\gamma$ is given simply by the size of the hole. For
finite holes, however, it is in general a complicated function of its size, shape and
position (see the many references in \cite{leaking}). For example, the decay rate can be
reduced if the hole contains a short-periodic orbit, because this implies that it
overlaps considerably with its images, reducing the amount of points that escape after
one step (this was observed for example in \cite{Paar,Ibere,Buni,transient1}).

A formula for $\gamma$ is available in terms of the equilibrium measure \cite{PY}: \be
\gamma=-\ln(1-\mu_e(h)),\ee where $\mu_e(h)$ is the $\mu_e$-measure of the hole $h$. This
can be understood as follows: for long times any initial probability measure converges to
$\mu_e$, and $\mu_e(h)$ gives the proportion of points which escape at one time step,
which must be equal to $1-e^{-\gamma}$. It is also possible to compute $\gamma$ using a
sum over periodic orbits contained in the hole \cite{Dorfman,Ott}.

One important characteristic of a fractal set is its dimension. Because $K_-$ is a simple
line in one direction, its total dimension is given by \be d(K_-)=1+\partial(K_-),\ee
where $\partial$ is called the {\em partial} fractal dimension. If the dynamics has
time-reversal symmetry, then $K_+$ has exactly the same dimension as $K_-$. Their
intersection, $K_0$, is locally equal to the product of their fractal parts and thus has
\be d(K_0)=2\partial(K_-)\ee as its dimension (for systems without time-reversal
symmetry, the dimension of $K_0$ is the sum of the partial dimensions of $K_-$ and
$K_+$).

We only consider fractal sets ($K_-$, $K_+$ and $K_0$) characterized by a single fractal
dimension (i.e. their Minkowski and Hausdorff dimensions coincide). For fractal measures,
on the other hand, different notions of dimensions coexist. The most used ones are the
Minkowski (or box-counting) dimension $d_0$, the information dimension $d_1$ and the
correlation dimension $d_2$. These are in principle distinct but in most practical
situations they happen to be quite close to each other. It is known that they can be
generalized to a continuous function $d_\beta$ which is non-increasing (the so-called
R\'enyi dimensions), so $d_0\geq d_1\geq d_2$.

We are only interested in the natural measure $\mu_e$. Since it is supported on $K_-$,
its Minkowski dimension is given by $d_0(\mu_e)=d(K_-)$. Since it is continuous in the
direction of $K_-$, we can write, for every $\beta$, the corresponding dimension in terms
of a partial component: \be d_\beta(\mu_e)=1+\partial_\beta(\mu_e).\ee We will use $d$
without a subscript for the fractal sets and with a subscript when we refer to dimensions
of the natural measure. It is quite common, however, to talk about $d_\beta(K_-)$ when
one means $d_\beta(\mu_e)$, because of the uniqueness of the natural measure. It is well
known that there exists an important equation (the Kantz-Grassberger relation
\cite{Kantz}) involving the decay rate, the Lyapounov exponent and the partial
information dimension of $\mu_e$. This is \be\label{KG}
\partial_1(\mu_e)=1-\frac{\gamma}{\lambda}.\ee

Coming back to the open tribaker map, if we fill phase space uniformly with initial
conditions, then exactly one third of points will hit the hole at each step, leading to
$\gamma=-\ln(2/3)=\ln3-\ln2$. The partial dimension of $K_-$ is the fractal dimension of
the Cantor set, well known to be $\partial(K_-)=\ln2/\ln3$. The natural measure is
constant on $K_-$ and, therefore, all its fractal dimensions are the same,
$\partial_\beta(\mu_e)=\partial(K_-)$. This is not a general feature; it comes from the
fact that the stretching rate is constant for this system. Notice that the Lyapounov
exponent is $\lambda=\ln3$, so that the relation (\ref{KG}) is indeed verified.

In Section 7 we shall discuss maps for which escape is not ballistic, i.e. when a point
falls in the hole, it has a certain probability of escaping but can also continue being
propagated. A probability density is then only partially attenuated when coming in
contact with the hole, instead of being drastically cut. This is supposed to model, for
instance, refractive escape of light rays from a dielectric sample.

\section{Quantum maps}

Maps have long been used as toy models in classical mechanics, because they allow for
chaotic conservative dynamics even in a two-dimensional phase space. But they have also
enjoyed popularity in the context of quantum mechanics, the reason being that their
numerical implementation is quite simple. When a map is defined on a compact phase space,
for example, its quantum propagator is automatically finite dimensional.

Unlike for Hamiltonian systems, there is no standard procedure to quantize a map. The
general idea is that, given a conservative classical map $\M$, one should associate with
it a unitary operator $U$ (the propagator) acting on some Hilbert space. Naturally, some
kind of classical limit must be defined, in such a way that the dynamics of $\M$ is
recovered.

Since we are considering only the torus as our phase space, we must enforce periodicity
both in $q$ and in $p$. This implies that both directions become discretized, i.e. there
will be a finite number $N$ of `position' eigenstates $|q_n\rangle$ and `momentum'
eigenstates $|p_m\rangle$. Quantization of classical periodicity can be achieved with a
pair of arbitrary phases, so that \be \langle q_n+1|\psi\rangle=\langle q_n|\psi\rangle
e^{2\pi i\chi_q}, \quad \langle p_n+1|\psi\rangle=\langle p_n|\psi\rangle e^{2\pi
i\chi_p}.\ee For given values of $n$ and $m$, the corresponding coordinates on the torus
are \be q_n=\frac{n+\chi_p}{N},\quad p_m=\frac{m+\chi_q}{N}.\ee The case
$\chi_q=\chi_p=0$ corresponds to periodic boundary conditions, while $\chi_q=\chi_p=1/2$
correspond to antiperiodic boundary conditions. Position and momentum must be Fourier
related, \be \langle
p_m|q_n\rangle=\left(F^{\chi_q,\chi_p}_{N}\right)_{nm}=\frac{1}{\sqrt{N}}e^{-2\pi i
(n+\chi_p)(m+\chi_q)/N}.\ee Here both $n$ and $m$ range from $0$ to $N-1$. The dimension
$N$ takes the role of inverse Planck's constant: \be N=\frac{1}{2\pi \hbar}.\ee

Quantization of the dynamics consists in the construction of a quantum propagator $U$,
which is a $N$-dimensional unitary matrix responsible for taking one wavefunction to
another: \be \psi'(q_n)=\sum_{k=0}^{N-1}U_{nk}\psi(q_k).\ee This operator can always be
diagonalized, and its spectrum lies in the unit circle in the complex plane, namely all
$N$ eigenvalues are of the form $e^{i\theta_n}$ with $0\leq \theta_n<2\pi$.

The matrix $U$ must somehow reproduce the classical dynamics in the limit $N\to\infty$.
To visualize this limit one usually resorts to tools such as coherent states and Husimi
functions. In usual quantum mechanics, coherent states $|q,p\rangle$ are ground states of
harmonic oscillators, i.e. Gaussian wavepackets with minimum uncertainty. When realized
in the torus, they must incorporate periodicity (a recent review about coherent states
can be found in \cite{gazeau}). Therefore, they are defined in position representation as
\be \langle q_n|q,p\rangle=C\sum_{k=-\infty}^\infty e^{-2\pi ik\chi_q}e^{-\pi
N(q_n-q+k)^2}e^{2\pi iN(q_n-q/2+k)p},\ee where $C$ is a normalization constant.

The Husimi function $\mathcal{H}_\psi(q,p)=|\langle\psi|q,p\rangle|^2$ of a quantum state
$|\psi\rangle$ is a real, non-negative function defined over all phase space. It can thus
be normalized and interpreted as a probability density (however, it does not have the
correct marginals, e.g. $\int\mathcal{H}_\psi(q,p)dp\neq |\psi(q)|^2$). The Husimi
function of a coherent state, $\mathcal{H}_{|Q,P\rangle}(q,p)$, is a periodized Gaussian
in both directions, centered at the point $q=Q,p=P$, with widths $\sim 1/\sqrt{N}$.

For large $N$, the action of the quantum propagator on coherent states must satisfy the
semiclassical condition $U|q,p\rangle\approx|\mathcal{M}(q,p)\rangle$, i.e. it must take
the point $q,p$ to its image under the classical map. Actually, this evolution will
always introduce a stretching of the wavepacket along the unstable direction, coming from
the Lyapounov exponent. It is more correct to say that the action of the quantum
propagator must correspond, in the semiclassical limit, to the action of the
Perron-Frobenius operator: it takes an initial probability distribution centered at
$(q,p)$ (the Husimi function of the state $|q,p\rangle$) to another probability
distribution which is centered at $\mathcal{M}(q,p)$ (but is not the Husimi function of
the state $|\mathcal{M}(q,p)\rangle$).

This semiclassical approximation cannot hold for long times due to interference effects.
The similarity between the action of $U^t$ on the quantum side and $\PF^t$ on the
classical side must degrade with time $t$. For chaotic systems, it degrades exponentially
fast. Heuristically, it will break down at the time scale given by the time it takes a
minimum wavepacket (of width $\sim 1/\sqrt{N}$) to stretch to the size of the system.
This is called the Ehrenfest time and, since the stretching is regulated by the Lyapounov
exponent according to $e^{\lambda t}$, it is given by $\frac{1}{2\lambda}\ln N$.

Notice that other definitions of Ehrenfest time are possible. For example, the time it
takes to stretch a state which is squeezed along the stable manifold into a state which
is squeezed along the unstable one is $\frac{1}{\lambda}\ln N$. This is equivalent to
stretching the smallest length, $1/N$, to the size of the system and has also been called
the Ehrenfest time in the literature (indeed this is the time when some interference
effects start to become important, see for example \cite{Saraceno1}). Yet another notion
of Ehrenfest time, related to escape through holes, will be presented below. The
quantum-classical correspondence for open chaotic systems, with special emphasis on
different Ehrenfest times, was reviewed in \cite{openreview}.

Let us now mention, without any technical details, some of the quantizations that have
been devised for the classical maps discussed in the previous section.

The first quantization of the baker map, introduced in \cite{balasz}, postulated the
following propagator: \be U=\left(F_{N}^{0,0}\right)^{-1}\left(
\begin{array}{cc} F_{N/2}^{0,0} & 0 \\0 &  F_{N/2}^{0,0} \\\end{array}
\right).\ee It is compatible with periodic boundary conditions and can only be defined on
even-dimensional spaces, reflecting the nature of the underlying map. It can be verified
that this operator has the baker map as its classical limit, but an inversion symmetry of
that map does not hold exactly. Later \cite{Saraceno1}, another quantization with
antiperiodic boundary conditions was suggested, that had exact symmetries. In this
quantization $F_{N}^{0,0}$ is replaced by $F^{1/2,1/2}_{N}$.

The quantization of cat maps was considered in \cite{Hannay,cat1,cat2}. Here the fact
that the classical action is quadratic was used to define a semiclassical quantization.
For the particular cat map (\ref{cat}) we mentioned previously, this results in a matrix
$U$ whose elements are quite simple: \be U_{nm}=\sqrt{\frac{i}{N}}e^{2\pi
i(n^2-nm+m^2)/N},\ee with $n,m$ between $0$ and $N-1$.

Finally, the standard map can also be quantized \cite{Casati,Izrailev}. It can be
realized as a kicked rotator, which leads to a quantization in terms of the product of
free propagation and kicking. More convenient is the following closed formula in position
representation: \be \fl U_{nm}=\sqrt{\frac{i}{N}}e^{\frac{i\pi}{N}(n-m)^2}\exp\left\{-
\frac{iNK}{4\pi}\left[\cos^2(2\pi n/N)+\cos^2(2\pi m/N)\right]\right\}.\ee This actually
corresponds to a version of the system which displays time-reversal symmetry.

A few experimental realizations have been suggested for the quantum maps discussed above.
In \cite{opticalbaker} an optical set-up of the baker map was proposed, while a possible
realization using nuclear magnetic resonance was discussed in \cite{nmrbaker}. Already in
\cite{Hannay} the possibility of realizing the quantum cat map using Fresnel diffraction
by a periodic grating was suggested. Finally, the quantum kicked rotator has already been
realized in the laboratory by placing cold atoms in a pulsed standing wave
\cite{kickedexp1,kickedexp2}.

It is natural to expect that in the semiclassical limit the Husimi functions of
eigenstates of $U$ should converge, in some weak sense, to invariant probability
distributions. As we have seen, these are either related to $f_e$ or singular. It is
known that, as $N\to\infty$, almost all eigenstates will become equidistributed, i.e.
their Husimi functions will converge to $f_e$. In other words, as one diagonalizes $U$ in
larger and larger dimensions, the fraction of equidistributed states should grow to
$100\%$. This property is called quantum ergoditicy. A stronger property is that of
quantum unique ergodicity. This property holds when we can say that all quantum states
become equidistributed without exception, which is a stronger statement than saying that
their fraction becomes $100\%$. Quantum ergodicity and quantum unique ergodicity have
been extensively investigated for cat maps and baker maps (see, for example,
\cite{QE1,QE2,QE3,QE4}). A recent review appears in \cite{anatomy}.

Numerically, it is easy to find for finite $N$ some eigenstates with Husimi functions
concentrated on periodic orbits. This phenomenon has been termed `scarring'. For
continuous-time systems, there is a large literature on this subject starting with
\cite{heller}; for maps see \cite{closedscars1,closedscars2} and also \cite{anatomy}.

In order to open a quantum map, what we have to do is identify a sector of the Hilbert
space with the hole. This is done by `projecting out' of it, i.e. the open propagator is
\be \widetilde{U}= U\Pi,\ee where $U$ is the closed propagator and $\Pi$ is the projector
on the complement of the hole. It is most common to use a hole that is a strip of width
$\delta$ parallel to the momentum axis. In that case $\Pi$ will be diagonal in position
representation. The value of $\delta$ sets a new length scale in the problem
\cite{openreview}. We can define the escape Ehrenfest time $\tau_e$ as the time it takes
to stretch the fundamental size $1/N$ up to $\delta$. This satisfies $e^{\lambda
\tau_e}=N\delta$. If we let $K$ be the number of quantum states that fit in the hole,
i.e. the dimension of the kernel of the projector $\Pi$, then clearly
$\tau_e=\frac{1}{\lambda}\ln K$.

The eigenvalues of $\widetilde{U}$ will be located inside the unit disk in the complex
plane. They can be written as \be z_n=e^{i\theta_n-\Gamma_n/2},\ee so that $\Gamma_n$ can
be interpreted as a decay rate (it corresponds to what was $\Gamma/\hbar$ in the more
general Hamiltonian setting). The set of all $N$ eigenvalues comprises the resonance
spectrum of this `scattering' problem. Resonances in the vicinity of the origin as seen
as short-lived, while any finite value of $\Gamma_n$ can be understood as being
long-lived.

\section{Fractal Weyl law}

The Weyl law is a very basic result in the asymptotics of energy levels for closed
systems. In a nutshell, it says that each stationary quantum state occupies a volume
$(2\pi\hbar)^\nu$ in phase space, where $\nu$ is the number of degrees of freedom
\cite{brack}. This means that, to leading order in $\hbar$, the number of states with
energy less than $E$ is proportional to $\hbar^{-\nu}$. Alternatively, let $\Omega(E)$ be
the set of phase space points with energy $E$. It is obvious that this set has dimension
$d_0=2\nu-1$. The density of states around $E$ is then proportional to the size of
$\Omega(E)$ and scales with $\hbar$ as $\hbar^{-(d_0+1)/2}$.

For open systems, a similar idea can be applied to the resonance spectrum. One can choose
a small but fixed region of the complex plane, around a point with real part $E$ and
imaginary part $-\hbar\Gamma/2$ (such that the decay rate is fixed), and ask how the
number of resonances inside it grows as $\hbar\to\infty$. For continuous-time chaotic
flows, it is conjectured that this number should behave like $\hbar^{-(d_0+1)/2}$, where
now $d_0$ is the fractal dimension of the classical chaotic saddle at energy $E$. This is
known as the fractal Weyl law (FLW). In the context of continuous time, it has been
checked for scattering from three gaussian `bumps' \cite{lin1,lin2}, three rigid disks
\cite{prl91wl2003}, four rigid spheres in 3D \cite{spheres} and a modified H\'enon-Heiles
potential \cite{borondo}. The only rigorous result available to date is that the exact
number of resonances is bounded from above by the FWL \cite{boundsFWL1,boundsFWL2}.

\begin{figure}[t]
\includegraphics[scale=0.9]{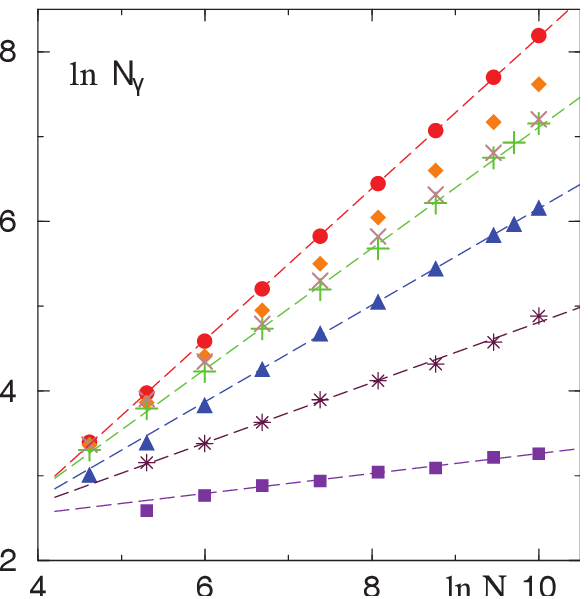}\hspace{0.5cm}\includegraphics[]{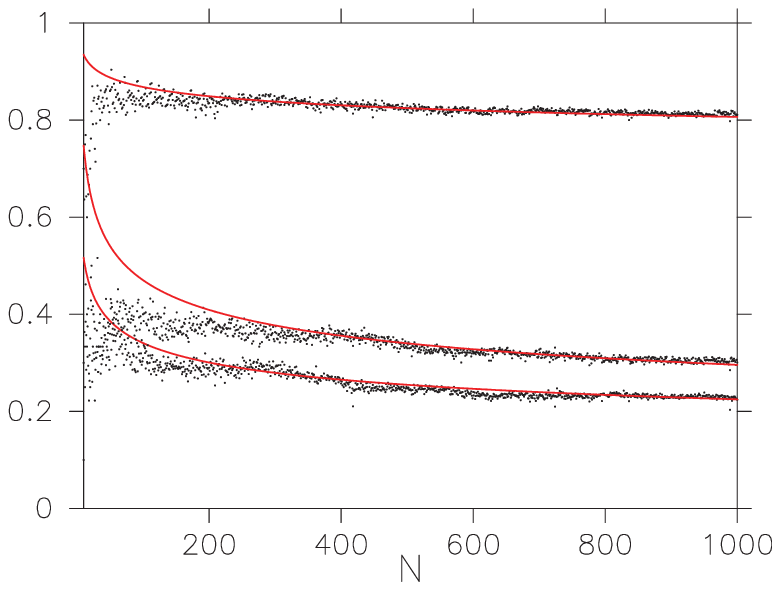}
\caption{(color online) Numerical verification of the fractal Weyl law. In the left panel
[Figure reprinted with permission from \cite{shepe}; Copyright (2008) by the American
Physical Society], the logarithm of the number of states having decay rate smaller than
some constant value is plotted against $\ln N$, for the kicked rotator with different
holes. In the right panel [Figure reprinted with permission from \cite{scar0}; Copyright
(2008) by the American Physical Society], the fraction of states having decay rate
smaller than some constant value is plotted against $N$, for the cat map with different
holes. In all cases, the FWL prediction (lines) is in excellent agreement.}
\end{figure}

\begin{figure}[t]
\includegraphics[]{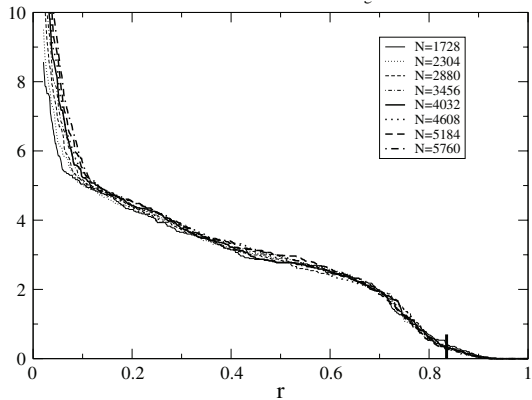}
\caption{Numerical verification of the fractal Weyl law for the baker map. The quantity
$n(r,N)/N^{-d/2}$ is plotted against $r$, where $n(r,N)$ is the number of eigenstates of
the map with modulus smaller than $r$ and $d$ is the fractal dimension of $K_0$. The plot
shows that the results for many values of $N$ lie on top of each other, validating the
scaling. From \cite{NonRub}, reprinted with permission.}
\end{figure}

A vague heuristic justification of the FWL can be presented as follows (a more detailed
account is given in \cite{lin1}). States with energy $E-i\Gamma/2$ have lifetime
$\hbar/\Gamma$. From basic principles, these states should be related to structures in
classical phase space which remain inside the system for that long. By analogy with the
usual Weyl law, their number is estimated as being proportional to the volume of a
`coarse-grained' version of the saddle, of thickness proportional to $\hbar$. As
$\hbar\to 0$, this volume is related to the fractal dimension.

The FWL takes on a particularly simple form for quantum maps, because of dimensional
reduction, no energy-dependence of the dynamics and the fact that the total number of
resonances is equal to $N$. In that case the conjecture is that the number of resonances
with modulus larger than some $r$ should grow like $C(r)N^{d/2}$, where $C(r)$ is some
unknown function of $r$ and $d$ is the fractal dimension of the saddle $K_0$ discussed in
Section 2. Notice that \be \frac{d(K_0)}{2}=d(K_-)-1=\partial(K_-),\ee where $d(K_-)$ is
the dimension of the backward trapped set and $\partial(K_-)$ is the associated partial
dimension. In \cite{NZ1,NZ2} the FWL was verified numerically for the open triadic baker
map and rigorously proven for a non-canonical quantization of this map. It has also been
verified for the kicked rotator \cite{henning,shepe} and for the cat map \cite{scar0}.
Again, only upper bounds can be rigorously proven \cite{bounds}. In Figures 3 and 4 we
show numerical verification of the FWL for three types of open quantum maps.

What is the function $C(r)$? The first point that needs to be settled is whether this
function is universal or system-dependent. For the kicked rotator \cite{henning} it is
quite close to the universal prediction from random matrix theory (see Section 8), but
this is not true for the baker map \cite{NZ1}. In fact, in this latter case there seems
to be a gap in the middle of the spectrum, for which no explanation has been offered so
far. If the function $C(r)$ is system-dependent, it remains to be determined what kind of
classical information goes into it.

One theoretical approach to the FWL, put forth in \cite{henning}, is via the construction
of short-lived states. We sketch its argument. Let $K$ be the dimension of the kernel of
$\widetilde{U}$ (the number of quantum states that fit in the hole). Let $\lambda$ and
$\gamma$ be the Lyapounov exponent and the escape rate of the system and let
$\tau_e=\frac{1}{\lambda}\ln K$ be the (escape) Ehrenfest time. Remember that
$\mathcal{R}^{-m}$ is the set of points that escape in exactly $m$ steps. A minimal
wavepacket $|\psi\rangle$ supported in $\mathcal{R}^{-m}$ will therefore reach the hole
in $m$ steps. If $m$ is smaller than $\tau_e$, this wavepacket will not have spread too
much and will still fit in the hole. It will then be almost completely annihilated and we
will have $\widetilde{U}^m|\psi\rangle\approx 0$, i.e. it will be a generalized
eigenvector of $\widetilde{U}$ with zero eigenvalue. This process is illustrated in
Figure 5.

What we mean by saying that $|\psi\rangle$ is a generalized eigenvector is that
$|\psi\rangle$ and its images under $\widetilde{U}$ form a Jordan chain. In other words,
the fractal Weyl law is related to the difference between the algebraic and the geometric
multiplicity of the null eigenvalue. These ideas were further discussed in \cite{kopp},
where it was suggested that, instead of being diagonalized, the open map should be
subject to a Schur decomposition, $\widetilde{U}=QTQ^\dag$, where $Q$ is unitary and $T$
is upper triangular. This has the advantage that, unlike the eigenvectors, the columns of
$Q$ are orthogonal.

\begin{figure}[t]
\includegraphics[scale=1.3]{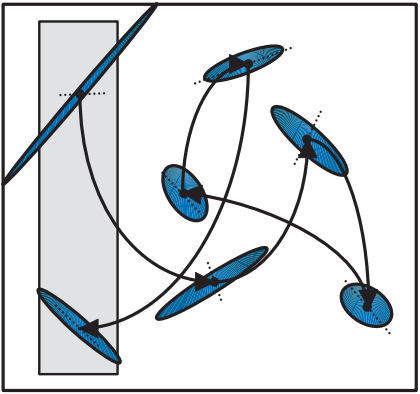}
\caption{(color online) Illustration of how propagation up to Ehrenfest time can lead to
almost perfect escape of a wave packet. The initial packet, whose center lies in
$\mathcal{R}^{-6}$, is very stretched along the stable manifold of the dynamics. Upon
propagation, it becomes stretched along the unstable direction, but still localized well
enough to fit in the hole. It is therefore an approximate generalized eigenstate of the
open map with vanishingly small eigenvalue. From \cite{openreview}, reprinted with
permission.}
\end{figure}

In essence, this identifies the states comprising the short-lived sector of the resonance
spectrum. To estimate their fraction, we need only estimate the total area of all
$\mathcal{R}^{-m}$ with $m$ up to $\tau_e$. Since the fraction of points remaining after
$t$ steps is $e^{-\gamma t}$, that area behaves like $\int_0^{\tau_e}e^{-\gamma t}dt\sim
1-e^{-\gamma \tau_e}$. Therefore, the short-lived sector amounts to a fraction of the
spectrum that scales as $N(1-K^{-\gamma/\lambda})$. Since $K$ is proportional to $N$,
this implies that the total number of states in the long-lived sector, which is counted
by the fractal Weyl law, should scale as $N^{\partial_1}$ where
$\partial_1=1-\gamma/\lambda$ is the partial information dimension of the natural measure
$\mu_e$.

The above arguments provide some understanding of the physical origins of the FWL, but
the exponent they predict is $\partial_1(\mu_e)$ instead of $\partial_0(\mu_e)$. Rigorous
upper bounds involve $\partial_0$. Still, the information dimension was also used in
\cite{shepe}, where the kicked rotator is studied in a range of decay rates, and again in
\cite{scar0} for the cat map with several different holes. In both these works very good
numerical agreement was obtained. On the other hand, it was shown in \cite{NonRub} that
the true exponent really is closer to $\partial_0$ by explicit construction of a system
where $\partial_0$ and $\partial_1$ are very different. In general, however, these two
dimensions are quite close; large deviations can only be expected for systems where the
local stretching factor is very far from being constant (which is the case in
\cite{NonRub}).

Another approach to the FWL was developed in \cite{scar1} and \cite{scar2}. This is based
on semiclassical approximations for the long-lived states, and the basic idea is that
these should be related to periodic points, which belong to the saddle. An approximate
basis for the long-lived sector is constructed by building wavepackets concentrated on
periodic orbits of period $T$, with $T$ up to $\tau_E=\frac{1}{\lambda}\ln N$. These
wavepackets are then propagated with the open quantum map, thereby acquiring some
information about short-time dynamics and providing an improved basis, the so-called scar
functions. Since the number of points with period less or equal to $\tau_E$ grows like
$e^{(\lambda-\gamma) \tau_E}$ (the quantity $\lambda-\gamma$ is the Kolmogorov-Sinai
entropy), the dimension of the basis is proportional to $N^{\partial_1}$. A matrix of
this dimension is then introduced as an approximation of the quantum propagator in the
long-lived sector. Its spectrum is in good agreement with exact results, validating the
approximation. This is shown for the baker map in Figure 6. Notice the presence of a gap
in the bulk of the spectrum.

This semiclassical approximation was critically discussed in \cite{moretrans}, where it
is argued that it reproduces the spectrum, but in order to correctly describe quantum
resonance wave functions semiclassically it is necessary to take into account more
information than only the chaotic saddle. The authors of \cite{moretrans} speculate that
diffraction effects and/or classical information outside the saddle should be important
and must be incorporated. This is yet to be systematically investigated. Another point
raised in \cite{moretrans} is that perhaps in practice the fractal Weyl law only holds
once $\hbar$ becomes much smaller than the size of the hole, otherwise it may be affected
by diffraction.

In \cite{bounds} a rigorous theory was developed to the construction, based on an open
quantum map of dimension $N$, of an auxiliary operator whose rank is $\sim
N^{\partial_0}$ and whose spectrum reproduces the exact resonances. That operator is
conceptually similar to the one based on short periodic orbits that was discussed above.

\begin{figure}[t]
\includegraphics[scale=0.9]{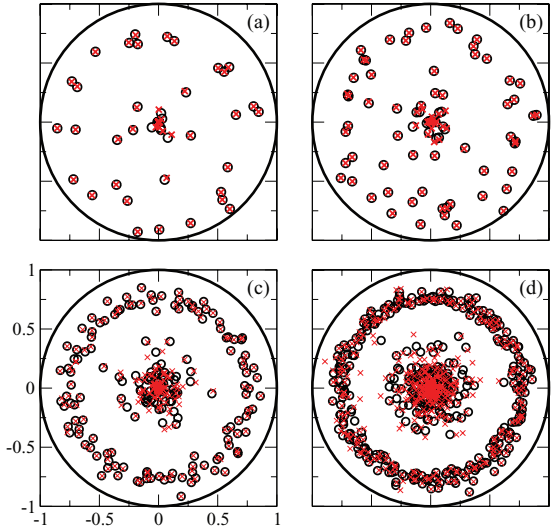}
\caption{(color online) Exact spectrum of the open triadic baker map (circles) and the
approximation to the long-lived sector obtained using scar functions (crosses). Four
values of $N$ are shown: 81, 177, 597, 1821. In each case, the approximation uses a basis
of size proportional to $N^{-d/2}$ where $d$ is the dimension of $K_0$. Good agreement is
obtained, validating the FWL scaling. The internal gap has not been explained. Figure
reprinted with permission from \cite{scar2}. Copyright (2012) by the American Physical
Society.}
\end{figure}

Notice that the two approaches we just delineated are different and somehow complementary
to each other. While \cite{henning} arrives at the FWL by estimating the size of the
short-lived sector, associated with regions of fast escape, \cite{scar1,scar2} attempt to
estimate the size of the long-lived sector, expected to be related to periodic orbits. In
both cases some notion of Ehrenfest time plays an important role.

The Weyl law for systems whose dynamics is a mixture of chaotic and regular regions was
discussed in \cite{kopp}. It was found that stability islands led to resonances with very
small decay rate, approximately following a usual Weyl law, while states with
intermediate decay rates obeyed a Weyl law with a fractional exponent; however, this
exponent was not related to the dimension of any fractal set. The subject was also
investigated in \cite{shudo} for a system with sharply divided phase space. States
associated with sticky motion at the border of the stability island satisfied Weyl laws
with fractional exponents, related to the power law behaviour of the classical survival
probability.

\section{Resonance gap}

The fractal Weyl law is a prediction about how the number of long--lived resonances
scales with $\hbar$ or, for quantum maps, with the dimension $N$. A different type of
question is how these resonances are distributed in the complex plane. For maps, we
denote them by $z_n=e^{i\theta_n}e^{-\Gamma_n/2}$, so that $|z_n|^2=e^{-\Gamma_n}$. In
particular, it is natural to focus interest on the longest-lived ones because they
usually leave clear signatures on scattering signals. Numerical experiments reveal that
the distribution of decay rates $\Gamma_n$ tends to have a maximum around the classical
decay rate $\gamma$, with a long tail for $\Gamma>\gamma$ and a rather short one for
$\Gamma<\gamma$. This has been observed for example in
\cite{shepe,ermann,Borgonovi,Shepe99}, and we show an example in Figure 7.

\begin{figure}[t]
\includegraphics[scale=1.1]{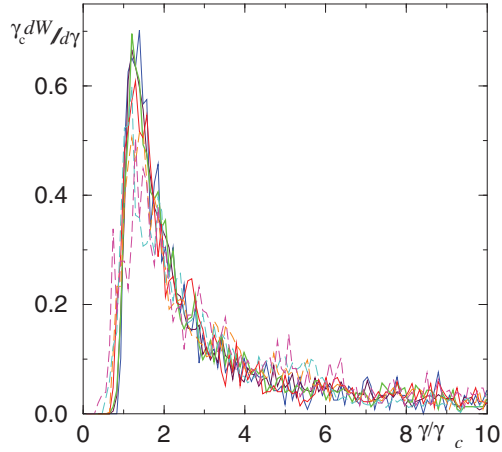}
\caption{(color online) Distribution of decay rates, scaled by the classical one
$\gamma_c$, of the open kicked rotator, for many different values of $N$. A limiting
behaviour as $N\to\infty$ and a sharp peak very close to $\gamma_c$ are clear. Figure
reprinted with permission from \cite{shepe}. Copyright (2008) by the American Physical
Society.}
\end{figure}

In a sense, states with $\Gamma<\gamma$ have anomalously slow decay. They were termed
`supersharp' resonances in \cite{ssr}. A natural question is whether there is a limit to
how sharp a resonance can be. Can the quantum decay rate be arbitrarily small? Evidence
is on the contrary. It is generally expected that supersharp resonances should become
less frequent in the semiclassical limit, i.e. that a true gap develops in the spectrum
as $N\to\infty$.

Already in \cite{gaspard} (see also \cite{ndisk}) it was found that a lower bound could
be proved for the decay rates of chaotic resonances in the limit $\hbar\to 0$. Its
formulation involves something called the topological pressure $P(\beta)$. This is
related to the leading eigenvalue of a generalization of the Perron-Frobenius operator
introduced by Ruelle. We do not discuss this so-called thermodynamic formalism here, and
instead refer the reader to \cite{Gaspard,Dorfman}. The topological pressure has some
important properties, namely its value at $1$ gives the decay rate, $P(1)=-\gamma$, its
derivative at this point gives the Lyapounov exponent, $P'(1)=-\lambda$, and it has a
zero at the partial dimension of the unstable manifold of the saddle,
$P(\partial(K_-))=0$. For the open triadic baker map it is given by
$P(\beta)=\ln2-\beta\ln3$; in general it is a convex decreasing function.

The available rigorous lower bound is that all $\Gamma_n$ must be larger than $-2P(1/2)$
as $N\to\infty$. This result has been revisited recently with more rigor in \cite{Gap}.
Interestingly, this bound is only effective if $\partial(K_-)<1/2$, because $P(\beta)$ is
a decreasing function with a zero at $\partial(K_-)$, so $P(1/2)>0$ if
$\partial(K_-)>1/2$. This would just predict that all $\Gamma_n$ are larger than a
negative value, something that is always true. In other words, only for saddles that have
low enough dimension, or are `filamentary' enough, can a true lower bound really be
established.

It is known that $2P(\beta)>P(2\beta)$ and hence $-2P(1/2)$ is always smaller than the
classical decay rate $\gamma=-P(1)$. Therefore, this lower bound still allows for quantum
decay that is slower than classical. At present, it is not clear whether in the limit
$N\to\infty$ one can still find resonances with decay rate arbitrarily close to
$-2P(1/2)$ or if they tend to be larger than $\gamma$. In other words, if a larger gap
can appear. An approach based on random matrix theory would predict the latter, as we
will see in Section 8.

Numerical experiments conducted for the baker map were somewhat inconclusive in this
respect \cite{NonRev}. In this system, the smallest decay rate $\Gamma_0$ seems to be
influenced by the presence of discontinuity lines and seems to be different for even/odd
eigenstates when the saddle intersects those lines. For the kicked rotator \cite{ssr}, on
the other hand, it was clearly observed that $e^{-\Gamma_0}-e^{-\gamma}$ approaches zero
as a power law $N^{-\delta}$. In fact, the exponent $\delta$ was found to be
approximately given by $\frac{d(K_0)}{2}-\alpha$, i.e. the exponent from the fractal Weyl
law minus a constant.

Also studied numerically in \cite{ssr} was the distribution of supersharp resonances. It
was found that they do not conform to the fractal Weyl law, i.e. their number does not
increase in proportion to the bulk of the spectrum. Instead it seems that are
approximately $N^\alpha$ such states, where $\alpha$ is the same constant as in the
previous paragraph. Whether this exponent is universal or system-specific and what
classical information it contains are open problems. Notice that therefore the number of
resonances with $\Gamma<\gamma$ grows with $N$, but at the same time the distance
$\Gamma_0-\gamma$ decreases with $N$. Similar results and conjectures have been surfaced
in the study of the Laplace resonance spectrum on hyperbolic surfaces of infinite volume
\cite{Naud1,Naud2}.

\begin{figure}[t]
\includegraphics[scale=0.9]{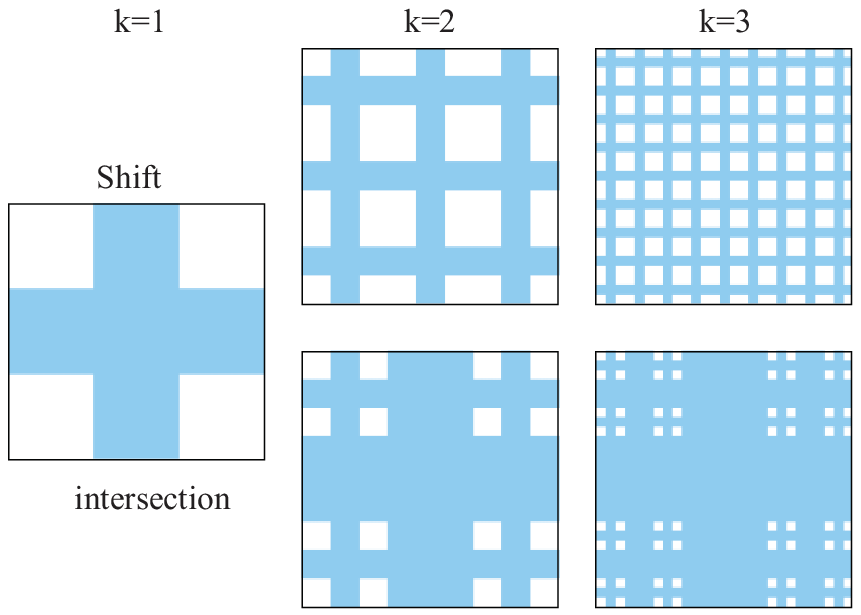}\hspace{0.5cm}\includegraphics[scale=0.9]{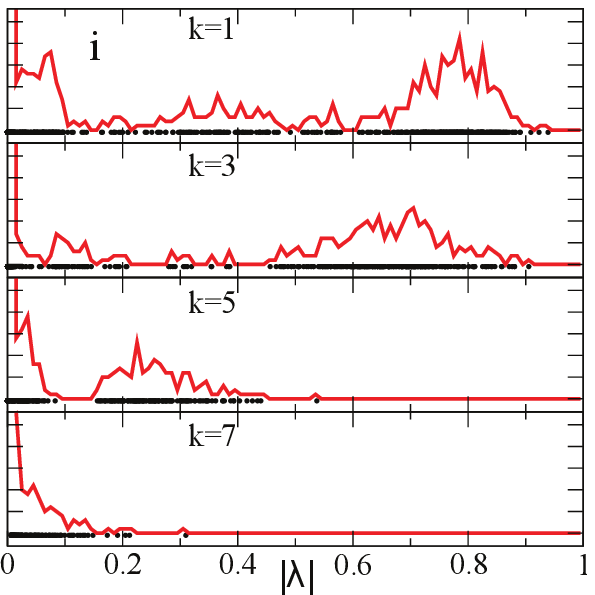}
\caption{(color online) Left: two families of open baker maps that have different holes
and different short-time dynamics, but the same decay rate and the same saddle. Right:
for the `intersection' family, the spectral radius decreases with $k$. Figure reprinted
with permission from \cite{transient2}. Copyright (2012) by the American Physical
Society.}
\end{figure}

The smallest decay rate, unlike the Weyl law exponent, seems to be very sensitive to the
short-time dynamics of the system, as observed in \cite{transient2}. In this paper two
families of baker maps were defined, along with their quantizations, the shift and the
intersection families, $B^{(s)}_k$ and $B^{(i)}_k$ for $k\geq 1$ (see left panel of
Figure 8). Inside each family, all members have the same Lyapounov exponent and decay
rate (and hence the same partial information dimension for their natural measures), but
very different short time dynamics. In the intersection family, members have holes that
grow with $k$, in such a way that only $(2/3)^{2k}$ of all initial conditions remain
after the first step. It was observed that the smallest decay rate grows with $k$, so
that all resonances have fast decay for large $k$ (see right panel of Figure 8). On the
other hand, in the shift family the area of the hole is independent of $k$ and so is
$\Gamma_0$.

\section{Eigenvectors}

As discussed in Section 3, open quantum propagators are not unitary matrices and their
left and right eigenvectors are different, \be
\widetilde{U}|\Psi_n^R\rangle=z_n|\Psi_n^R\rangle,\quad
\langle\Psi_n^L|\widetilde{U}=z_n\langle\Psi_n^L|.\ee Just like for closed systems, one
would expect on general grounds that in the semiclassical limit these wavefunctions
should be related to classical structures. A natural question is the analogue of the
quantum ergodicity problem: what are the possible semiclassical limits of Husimi
functions of resonance wavefunctions?

The issue of quantum ergodicity for open maps is actually much more difficult than for
closed ones, because there are many conditionally invariant measures for any given decay
rate, and it is not clear which ones would be more natural as semiclassical limits of
Husimi functions.

A few general results were obtained in \cite{knps}. For example, Husimi functions of
right eigenstates, $\mathcal{H}_{\psi_n^R}(q,p)=|\langle q,p|\Psi_n^R\rangle|^2$, become
concentrated, in the semiclassical limit, on the backward trapped set $K_-$. This is true
in the following sense: suppose a sequence of resonances with increasing $N$, such that
the corresponding eigenvalues converge to some value $z$ which is different from zero.
Then the sequence of (right) Husimi functions will converge to zero at points which do
not belong to $K_-$. We show two illustrative examples in Figure 9. Analogously, left
eigenstates concentrate on $K_+$ as $N\to\infty$. The results in \cite{knps} do not
indicate how fast the value of $\mathcal{H}_{\psi_n^R}(q,p)$ goes to zero if $(q,p)\notin
K_-$, but in \cite{NonRub} it was shown that this happens (at least for the baker map)
exponentially fast, i.e. like $e^{-cN}$ for some constant $c$.

\begin{figure}[t]
\includegraphics[scale=1.1]{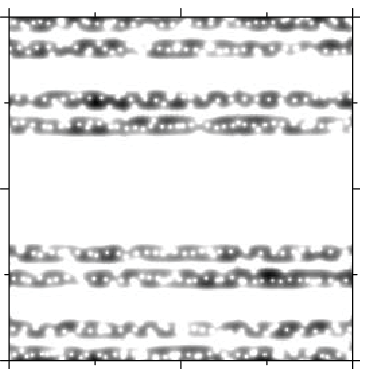}\hspace{0.5cm}\includegraphics[scale=1.1]{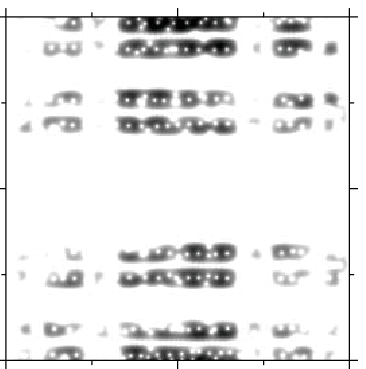}
\caption{Two Husimi function of right eigenstates of the open triadic baker map, for
$N=1500$. Intensity grows from light to dark. Eigenvalue modulus is $0.84$ for the left
one, which is approximately uniformly distributed along $K_-$, and $0.46$ for the right
one, which shows some localization on the hole and $\mathcal{R}^{-1}$. From
\cite{NonRev}, reprinted with permission.}
\end{figure}

\begin{figure}[t]
\includegraphics[scale=0.9]{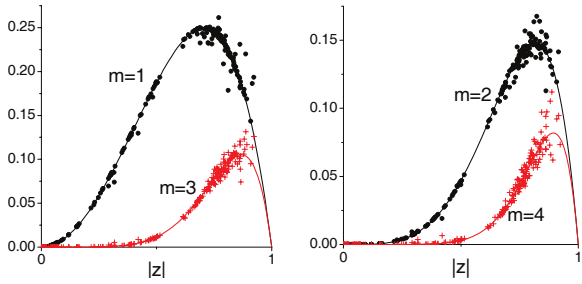}
\caption{(color online) The weight of Husimi functions of right eigenstates of the open
triadic baker map on the regions $\mathcal{R}^{-m}$, for $N=729$. The solid line is the
semiclassical approximation (\ref{weights}). Figure reprinted with permission from
\cite{knps}. Copyright (2006) by the American Physical Society.}
\end{figure}

It is also possible to obtain some semiclassical information about how these Husimi
functions are distributed on their support. First, remember that the sets
$\mathcal{R}^{-m}$ intersect the backward trapped set. Let these regions be
semiclassically quantized by phase space projectors $\pi_m$ (as discussed in
\cite{vallejos}). The open propagator thus satisfies
$\widetilde{U}^\dag\widetilde{U}=1-\pi_0$. It was shown in \cite{knps} that
\be\label{weights} \langle\Psi_n^R|\pi_m|\Psi_n^R\rangle\approx|z_n|^{2m}(1-|z_n|^2).\ee
The left hand side of the above equation measures the weight of
$\mathcal{H}_{\psi_n^R}(q,p)$ on the region $\mathcal{R}^{-m}$, and the right hand side
relates this to the decay rate. This relation is tested in Figure 10 for the baker map.

It was first noticed that the functions $\mathcal{H}_{\psi_n^R}(q,p)$ had fractal
signatures of the set $K_-$ in \cite{Shepe99}. In this paper it was suggested that, for a
sequence of states whose decay rate converges to the classical one, the (right) Husimi
functions should converge to the classical equilibrium measure. This would in fact be
consistent with the result (\ref{weights}) above, because since $\mu_e(h)=1-e^{-\gamma}$
and since the area decreases by $e^{-\gamma}$ at each step, we have \be
\mu_e(\mathcal{R}^{-m})=e^{-m\gamma}(1-e^{-\gamma}).\ee However, this convergence has not
been proven and numerical evidence is inconclusive. Equation (\ref{weights}) also
predicts that states with $\Gamma_n>\gamma$ must show some concentration on the hole and
its first pre-images and, conversely, that states with $\Gamma_n<\gamma$ (supersharp)
must avoid these sets and localize closer to the invariant set.

This problem was extensively discussed for the baker map in \cite{NonRub}. In particular,
this system admits a non-canonical quantization, known as the Walsh quantization, in
which it is exactly solvable (unfortunately it is not clear how general are the results
proven in this special setting). For example, right Husimi functions having self-similar
properties can be explicitly constructed for this system. This was discussed again in a
slightly different setting in \cite{knns} where, among other things, a kind of quantum
unique ergodicity was proven at the edges of the spectrum.

Related to the problem of quantum ergodicity is the phenomenon of quantum scarring. This
is a generic name given to concentration of quantum states in the vicinity of periodic
orbits. For open systems the question of scarring carries an extra interest, because of
the interplay with the decay rate. States with small decay rate must survive in the
system for a long time, and it is natural to expect them to be scarred along periodic
orbits. States with $\Gamma>\gamma$ can show scarring, and this has been observed. But,
as discussed in relation with (\ref{weights}), scarring is more likely for
$\Gamma<\gamma$ (this was also suggested in \cite{Borgonovi}). Moreover, how the effect
behaves in the semiclassical limit is rather unclear. This is connected to the subject of
the previous Section, because a gap may develop in the spectrum such that no states with
$\Gamma<\gamma$ survive the semiclassical limit.

It was observed in \cite{Shepe99} that resonances of the kicked rotator with small decay
rates were scarred. The subject was taken up again in \cite{scar0}, where the open cat
map was studied. A set of quantum states specially suited for studying scarring, the
so-called scar functions already available for closed maps \cite{scarfuncs1,scarfuncs2},
was adapted to the open setting in \cite{scar1,scar2}, where it was observed that the
baker map with low $N$ had some resonance wave functions that could be obtained as a
superposition of only two of these states. However, it was also suggested in
\cite{moretrans} that classical information outside the saddle, such as diffraction
effects, must somehow be incorporated if a semiclassical approach is to accurately
reproduce resonance wave functions.

In this context, a new phase space representation of resonance states was introduced in
\cite{carlo}, which is a generalization of the Husimi function. It is given by \be
h_n(q,p)=\frac{\langle q,p|\Psi_n^R\rangle\langle \Psi_n^L|q,p\rangle}{\langle
\Psi_n^L|\Psi_n^R\rangle}.\ee Since $\langle q,p|\Psi_n^R\rangle$ is localized on $K_-$
and $\langle \Psi_n^L|q,p\rangle$ is localized on $K_+$, the quantity $h_n(q,p)$ must be
localized on their intersection, the saddle, and may be useful for revealing scarring
effects. This was verified for the baker map. It was also noticed that, for finite $N$,
the states with large decay rates can have significant values of $h(q,p)$ outside of that
fractal set.

\begin{figure}[t]
\includegraphics[scale=0.4]{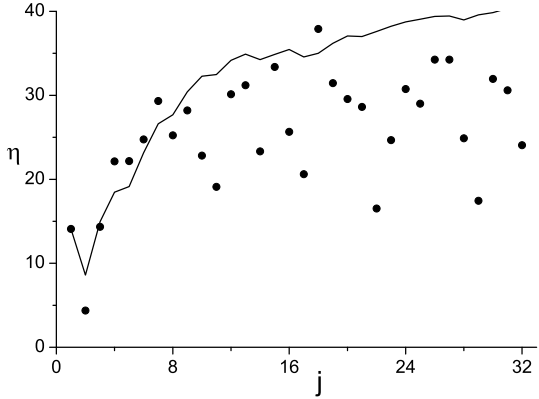}
\caption{The scarring estimator $\eta$ for the open triadic baker map at $N=243$. The
circles correspond to individual states, $\eta(h_j)$, ordered with increasing decay rate.
The line corresponds to the cumulative value $\eta\left(\sum_{i=1}^j h_i\right)$.
Courtesy of L. Ermann.}
\end{figure}

A quantity derived from  $h(q,p)$ was introduced in \cite{transient2} as a quantitative
measure of localization: \be
\eta(h)=\left(\frac{|h|_1/|h|_2}{|\rho|_1/|\rho|_2}\right)^2,\ee where \be
|h|_s=\left(\int |h(q,p)|^sdqdp\right)^{1/s}\ee and $\rho$ represents the density matrix
of any coherent state. As defined, $\eta$ is equal to $1$ for a coherent state (perfect
localization) and $N/2$ for a uniformly distributed state in a $N$-dimensional Hilbert
space. We see in Figure 11 that the value of $\eta$ behaves rather erratically as a
function of the decay rate for individual states. However, a cumulative value
$\eta\left(\sum_{i=1}^j h_i\right)$, where the sum runs over states with increasing decay
rate, shows an almost monotonic behaviour, with smaller decay rates corresponding to
stronger localization.

\section{Non-ideal escape}

A generalization of open systems which has attracted attention recently is the
possibility to define holes that are only partially transparent: when a point hits the
hole, it has a finite probability of escaping, but may also continue being propagated
inside. This may be called non-ideal or refractive escape, because it is supposed to
mimic the ray-splitting that takes place at the interface between two media with
different indices of refraction. It can also be used to model quantum dots coupled with
tunnel barriers or any kind of losses.

Let us suppose a dielectric sample of refractive index $n$, outside of which the
refractive index is $1$. When the wavelength is much smaller than the average radius of
the sample, we may consider propagation of rays. These follow straight lines, except when
they meet the boundary. According to Snell's law, they then experience specular
reflection. This is what is known as a dynamical billiard. It is usual to define boundary
coordinates as follows. Choose an arbitrary point $r_0$ at the boundary and let $q$ be
the distance along the boundary between a collision and $r_0$. Clearly, $0\leq q<L$,
where $L$ is the perimeter of the sample. The other coordinate, $p$, is the sine of the
angle between the incoming ray and the normal to the boundary at the collision point. It
is possible to show that in this coordinate system the dynamics is area-preserving.

The propagation of rays inside the dielectric sample is thus reduced to a discrete map in
the variables $(q,p)$, of the kind we have been considering. Now we must incorporate the
fact that, at each collision, some amount of light is refracted out of the sample. The
intensity of the reflected beam is given by Fresnel's law (for transverse magnetic
polarization, for instance): \be
R(p)=\left(\frac{\sqrt{1-p^2}-p_c\sqrt{1-(p/p_c)^2}}{\sqrt{1-p^2}+p_c\sqrt{1-(p/p_c)^2}}\right)^2\ee
and depends on the angle of incidence. Refraction does not occur if $|p|>p_c=1/n$, which
is called the critical angle.

Refraction can be modeled in a chaotic map by assigning an escape probability to each
point in phase space \cite{mapslaser}. An attempt to be realistic would try to simulate
Fresnel's law. More simply, one can define a finite hole corresponding to $|p|<1/n$ and
associate to it a constant escape probability. Now it is no longer possible to simply
follow initial conditions with time, because one must keep track of probabilities. This
can be solved by considering only the evolution of probability densities (this is also
discussed in \cite{leaking}).

We can start with the uniform density in phase space, for example, and let it evolve
under the partially open map. Equivalently, we can associate an initial intensity with
each point, and this intensity is reduced every time the point falls in the (partially
transparent) hole. In particular, the set of points whose intensity remains forever equal
to its initial value is nothing but the forward trapped set that would correspond to a
totally transparent hole. However, other regions may also have high intensity and be
important for long-time properties.

\begin{figure}[t]
\includegraphics[scale=1.1]{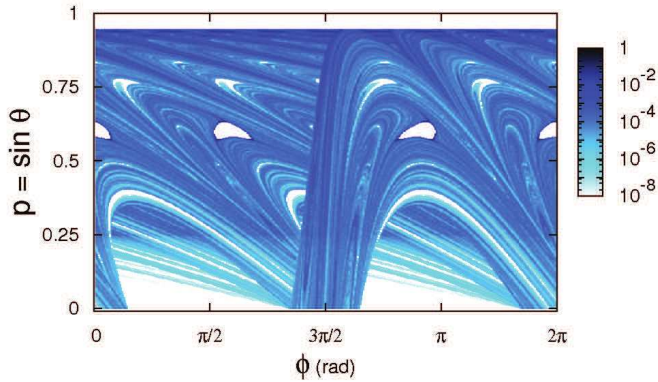}
\caption{(color online) The conditionally invariant natural measure of the dielectric
annular billiard, with refractive index equal to $4$, depicted in the boundary
coordinates described in the text (vertical axes is $p$, horizontal one is $q$).
Intensity grows from light to dark. Figure reprinted with permission from
\cite{altmann2009}. Copyright (2009) by the American Physical Society.}
\end{figure}

Strictly speaking, the system will have a (conditionally invariant) equilibrium measure
$\mu_e$ whose support (the `forward trapped set') is the entire phase space, since the
intensity of a point will never be exactly zero. In other words, $\mu_e$ will have a
trivial Minkowski dimension $d_0(\mu_e)=2$. However, the intensity after a long time will
be a wildly fluctuating quantity and usually there will exist non-trivial dimensions
$d_\beta(\mu_e)<2$ (naturally, the same is true for backward time evolution). Further
discussion of these points can be found in \cite{main,altmann2009,leaking}, along with
pictures of such intensities for the stadium, annular and cardioid  billiards,
respectively. We show an example in Figure 12.

Quantization of partially open maps poses no difficulty. We continue assuming that the
hole is a strip (or a union of strips) parallel to one of the axes. Instead of using a
projector $\Pi$, one needs only multiply the closed-system propagator $U$ by a diagonal
matrix whose elements are equal to $\sqrt{R}$ where the reflection probability is $R$ (in
principle, the phases of such elements are arbitrary; obviously, we assume $0\leq R\leq
1$).

The first question one would ask in this setting is about the analogue of the Weyl law.
Do we see a fractional exponent in the way the number of resonances scales with $N$? It
was initially suggested in \cite{main} that the Weyl law could be sensitive to the
different fractal dimensions present in the classical equilibrium measure. By counting
resonances in the dielectric stadium billiard, different exponents in the Weyl law were
found depending on the decay rate; moreover, these exponents were in the same range as
the numerically determined $d_\beta(\mu_e)$.

However, it was later proved \cite{NonPart,schenck} that in the semiclassical limit the
resonances tend to cluster at a particular value of the decay rate. More specifically,
let $R(q,p)$ denote the reflection probability at point $(q,p)$. In the long run, a
typical (ergodic) trajectory will sample all of phase space. At each point, its intensity
is multiplied by $R$. This process is additive in the logarithm, resulting in the average
decay $e^{-\langle \Gamma\rangle}$, where \be\label{avegamma} \langle \Gamma
\rangle=-2\int\log(R(q,p))dqdp.\ee It was shown in \cite{NonPart,schenck} that the number
of resonances with decay rate inside a fixed interval containing $\langle \Gamma\rangle$
scales with $N^1$ in the semiclassical limit. In other words, most resonances have this
decay rate. This result is illustrated in Figure 13. Since the resonances counted in
\cite{main} were in a range that contained $\langle \Gamma\rangle$, the fractional
exponents observed are probably due to insufficient statistics or the semiclassical
regime not been reached yet.

\begin{figure}[t]
\includegraphics[scale=1.3]{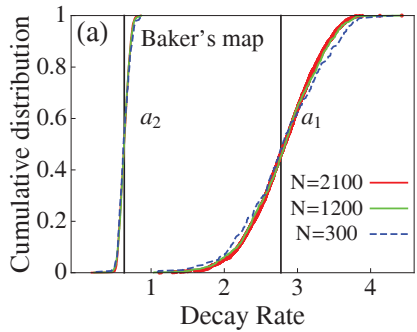}
\caption{(color online) Cumulative distribution function of decay rates for partially
open baker maps (curves $a_1$ and $a_2$ correspond to two different choices for hole
position and reflectivity). Vertical lines correspond to $\langle \Gamma\rangle$, and
clearly the distribution has an inflection point there. Notice how little change there is
with $N$. Figure reprinted with permission from \cite{NonPart}. Copyright (2008) by the
American Physical Society.}
\end{figure}

Another result presented in \cite{NonPart,schenck} is that, for strongly chaotic systems,
the width of the distribution of decay rates around $\langle \Gamma\rangle$ decays like
$(\ln N)^{-1}$. This was proven for a toy model and numerically observed (albeit not
quite clearly) for the baker map and a perturbed cat map. It is natural to expect that
this spectral width should depend on some type of variance of the function $R(q,p)$. In
particular, the width is zero if $R$ is constant all over phase space. This point has not
been investigated.

The distribution of resonances for partially open maps had already been considered
numerically in \cite{mapslaser} using the kicked rotator as a model. This paper also
suggested a random matrix theory approach (see next Section), which was later tested in
\cite{schomain} for the stadium billiard. Good agreement was obtained, as long as the
mean decay rate and the effective number of modes were suitably adjusted.

The phase space morphology of resonance states in the case of non-ideal escape has not
been investigated for maps, as far as this author is aware (see however
\cite{partial1,partial2,partial3}).

\section{Random matrix theory}

An approach which has been very successful in all areas of quantum chaos is the theory of
random matrices \cite{Mehta}. In essence, the specific relevant operator (Hamiltonian,
$S$ matrix, propagator) is abandoned and replaced by a random matrix, i.e. only
statistical properties are studied with respect to a certain ensemble of matrices. How
this ensemble is chosen is dictated by the situation at hand. Approaches based on random
matrices exist for several areas of physics, including the quantum mechanics of closed
chaotic systems \cite{haake} and chaotic transport \cite{beenakker}. Many examples can be
found in a recent Handbook \cite{handbook}.

To replace the Hamiltonian of a closed system, for example, it is important that
hermitian matrices be used. The simplest possibility is then to replace the matrix
elements with identically distributed Gaussian random variables, while respecting the
hermiticity constraint. This gives rise to the Gaussian Unitary Ensemble (it is called
Unitary because it is invariant under unitary transformations). If there is time-reversal
symmetry, the Hamiltonian can be made real without diagonalization. What is needed then
is an ensemble of real symmetric matrices, and this leads to the Gaussian Orthogonal
Ensemble. It is common practice to associate an index $\beta=2$ with complex matrices and
$\beta=1$ with real matrices (this is sometimes called the Dyson parameter). Gaussian
ensembles were recently reviewed in \cite{RevModPhys1}.

\subsection{Effective Hamiltonian approach}

It is possible to model scattering by writing the $S$ matrix as
$S(E)=1-iV^\dag\frac{1}{E-H_{eff}}V$, where $V$ is a $N\times K$ coupling matrix between
the system and the outside and $H_{eff}$ is an $N\times N$ effective non-hermitian
Hamiltonian. The poles of $S$ are then the eigenvalues of $H_{eff}$. This is sometimes
called the Heidelberg approach. Assuming energy-independence of the coupling elements and
neglecting direct processes, this latter matrix can be taken as
$H_{eff}=H-\frac{i}{2}VV^\dag$, and $H$ considered as being uniformly distributed in one
of the Gaussian ensembles. The reader can consult \cite{Fyod,FS,FSS,SavinFyod,RevModPhys}
for further information.

Different regimes can be studied depending on coupling strength and the interplay between
$K$ and $N$. The limit $N\to\infty$ is always intended. The number of channels $K$ may
remain fixed, in which case the hole is not classical in size. For small coupling, all
eigenvalues acquire small imaginary parts (smaller than the mean level spacing). For
large coupling, the phenomenon of resonance trapping occurs \cite{trap}: only a few
resonances acquire large imaginary parts, while most of them approach the real axis.

The case of classical opening, closer in spirit to the subject of this review,
corresponds to keeping the ratio $K/N$ fixed as $N\to\infty$. This corresponds to the
size of the hole, $K/N=h$. As we have seen, it makes sense to consider this as a small
variable. It can be shown \cite{haakeres,lehmann} that in this case the density of
resonances exhibits a gap: it vanishes for decay rates smaller than $h$, which can be
identified with the classical decay rate. Moreover, this density decays like $1/\Gamma^2$
for $\Gamma>h$.

\subsection{Approach via truncations}

The propagator of a closed quantum map is a $N$-dimensional unitary matrix, $UU^\dag=1$.
If there is time-reversal symmetry, this matrix is also symmetric. These are the only
constraints that must be imposed. The ensemble of random unitary matrices is nothing but
the unitary group equipped with its unique invariant (probability) measure, the Haar
measure. In this context, this ensemble is called the Circular Unitary Ensemble, the CUE.
If symmetry is imposed, $U^T=U$, we obtain the Circular Orthogonal Ensemble, the COE.

The propagator of an open quantum map is obtained by multiplying the closed propagator by
a projector. This is sometimes called a truncation: some columns of $U$ are set equal to
zero to produce $\widetilde{U}$. Therefore, it is natural to take $U$ as being uniformly
distributed in the CUE and introduce an ensemble of truncated unitary matrices (TCUE).
This was done in \cite{trunc}.

We therefore have $N$-dimensional unitary matrices truncated by multiplying from the
right by a projector. Let the kernel of the projector have dimension $K$, as before, and
let $M=N-K$ be the number of eigenvalues that are not identically zero. Just like for the
effective hamiltonian approach, in the `semiclassical' limit $N\to\infty$ there are two
choices for the behaviour of the truncation: the kernel dimension $K$ can remain fixed in
order to model `quantum' holes, or we can have $M=\mu N$ with fixed $\mu$ in order to
model `classical' holes.

Since any element of the TCUE is subunitary, the spectrum always lies inside the unit
disk in the complex plane. The joint probability density of the eigenvalues can be
exactly obtained. For large $N$, the original ensemble of $U$ is not relevant, i.e.
results are the same for both unitary and orthogonal classes. Apparently, the presence of
the hole is enough to effectively break symmetry. The probability distribution of the
modulus squared of the eigenvalues can also be found. Using the quantity $\mu$ mentioned
above, the distribution of $\rho=|z|^2=e^{-\Gamma}$ is given by \be\label{RMTdensity}
P(\rho)=\frac{1-\mu}{\mu}\frac{1}{(1-\rho)^2}\ee if $\rho<\mu$ and vanishes otherwise.

Since $\mu=1-h$ is the area of the complement of the hole, it is natural to interpret it
in terms of a fictitious decay rate as $\mu=e^{-\gamma}$. Therefore, this approach also
predicts a spectral gap in the limit $N\to\infty$, so that states with $\Gamma<\gamma$
become increasingly rare. Note however that the interpretation of $\Gamma$ as a decay
rate is not so straightforward for discrete time systems (maps), e.g. \be \int_0^\infty
e^{-\Gamma t/\hbar}dt=\frac{\hbar}{\Gamma},\quad\text{ while }\quad\sum_{n=0}^\infty
e^{-\Gamma n}=\frac{1}{1-e^{-\Gamma}}.\ee If we were to interpret $1-e^{-\Gamma}$ as a
decay rate, then according to (\ref{RMTdensity}) its density decays quadratically, in
agreement with the effective Hamiltonian approach.

In Figure 14 the prediction (\ref{RMTdensity}) is checked for the open kicked rotator
[not available in the Arxiv version]; good agreement is found, provided the dimension $N$
is renormalized according to the fractal Weyl law.


Notice that this approach is not able to give any insight into the fractal Weyl law
itself, since this is obviously a system-specific feature, i.e. it depends on details of
the dynamics. One way to see this is that although it is possible to build the escape
rate into random matrix theory, simply using the size of the hole, the same is not true
for the Lyapounov exponent (in a sense, this quantity is `infinite' in this theory).

Even though they are suppressed in the $N\to\infty$ limit, we can ask about statistics of
resonances inside the gap, the supersharp ones. For example, we can consider states with
$\rho=\mu+\epsilon/\eta$, with $\eta=\sqrt{N/2\mu(1-\mu)}$. These will approach the
maximum value from above in the semiclassical limit. It turns out that they have a finite
density function proportional to ${\rm erfc}(\epsilon)$, where erfc is the complementary
error function \cite{Khoruz}. Another interesting question is the distribution of the
smallest decay rate $\Gamma_0$. Equivalently, we can consider the largest eigenvalue
modulus squared, $\rho_0=e^{-\Gamma_0}$. It was shown in \cite{Rider,ssr} that the
modified variable $\eta^2(\rho_0-\mu)^2-\ln N+\ln\ln N$ satisfies a Gumbel distribution
as $N\to\infty$.

\subsection{Eigenvectors}

A different line of investigations would be to consider statistics of the eigenvectors.
Usually, the quantity \be K_n=\frac{\langle \Psi_n^L|\Psi_n^L\rangle\langle
\Psi_n^R|\Psi_n^R\rangle}{|\langle \Psi_n^L|\Psi_n^R\rangle|^2},\ee is considered (see,
for example, \cite{Patra,Mehlig}). This is called eigenvector correlator or Petermann
factor, and is related to the line width of the lasing mode in open resonators. Results
of this type were reviewed in \cite{FS}. This can be generalized to an off-diagonal
version $U_{nm}=\langle \Psi_n^L|\Psi_m^L\rangle\langle \Psi_n^R|\Psi_m^R\rangle$, which
was shown in \cite{YanSavin} to be related to the statistics of resonance width shifts
under external perturbations.

Another type of quantity, \be q_n^2=\frac{\sum_i ({\rm Im}\Psi_n^i)^2}{\sum_i ({\rm
Re}\Psi_n^i)^2},\ee where $\Psi_n^i$ is the $i$th component of a right eigenvector, was
considered in \cite{savin1,savin2}.

Research along this line has always been done within the effective Hamiltonian approach,
in the case when the number of states in the hole is much smaller than total number of
states, $K\ll N$. To the knowledge of this author, there are no works about statistics of
eigenvectors for classical holes or truncated unitary matrices.

\subsection{Non-ideal escape}

The case of non-ideal escape can also be treated in two ways, using effective
Hamiltonians or opening propagators. Within the first approach the finite transparencies
of the decay channels are incorporated into the coupling matrix $V$. When all channels
are equivalent the degree of resonance overlapping is controlled by the parameter $KT$
\cite{sokolov} where $T$ is the transparency and $K$ is the number of channels.

The propagator of partially open systems should be modeled as $\widetilde{U}=U\sqrt{G}$,
where $G$ is a fixed diagonal matrix with nonnegative entries $0\leq g_1\leq
g_2\leq\cdots\leq 1$ representing reflection probabilities and $U$ is uniformly
distributed in a circular ensemble. This was first considered in \cite{mapslaser}, with
$U$ in the COE, but only numerical results (for distribution of decay rates and
eigenvectors) were presented. Later, the exact density of states with $U$ in the CUE was
derived in \cite{Wei}. Its asymptotic behaviour as $N\to\infty$, obtained in
\cite{HL,Bogo}, is as follows: if $x=|z|$ denotes eigenvalue modulus, its cumulative
distribution $y(x)$ is implicitly defined by \be \Psi\left(\frac{y-1}{xy}\right)=y-1,\ee
where $\Psi$ is \be
\Psi(u)=\lim_{N\to\infty}\frac{1}{N}\sum_{j=1}^N\frac{ug_j}{1-ug_j}.\ee In particular, in
the semiclassical limit the eigenvalues of $\widetilde{U}$ lie inside an annulus in the
complex plane: their modulus squared must be larger than the arithmetic mean of the
eigenvalues of $G$ and smaller than their harmonic mean.

\section{Some open problems}

The topic of resonances is of wide interest, both theoretically and experimentally. There
is a variety of settings, motivations, applications, approaches, etc. that make it
impossible to provide a truly comprehensive review. We have focused on quantum maps, with
chaotic classical dynamics and classical openings. We hope to have touched upon some of
the more interesting points that have recently attracted attention and to have provided
some references for the interested reader.

We conclude this review with a few open problems. These are more general indications of
interest than specific questions. It is likely that as these matters are investigated
further other interesting problems will come to the front.

For systems with ballistic escape, we would like to mention the following points:
\begin{itemize}
\item Exponent in the fractal Weyl law. A rigorous proof of this, aside from the
non-generic Walsh quantization of the baker map, is still lacking. Heuristic
understanding has been achieved from different points of view, however, and such proof is
likely to be extremely technical.

\item Prefactor in the fractal Weyl law. Does the profile function of the
resonance spectrum $C(r)$ have anything to do with classical dynamics? The random matrix
theory prediction works well for the kicked rotator, but not for the baker map. Can this
be understood?

\end{itemize}

The next points apply to ballistic escape, but also make sense for systems with non-ideal
escape:
\begin{itemize}

\item Spectral gap. Under what conditions is it true that as $N\to\infty$ there are
    no resonances in a certain region? What exactly is the Weyl law for supersharp
    resonances?

\item Quantum ergodicity. What are the possible semiclassical limits for Husimi
functions with a given decay rate? Do they have self-similar properties? Is there a
preferred classical eigenmeasure?

\item Scarring. How does the scarring by periodic orbits behaves in the semiclassical
    limit? Is the amount of scarring related to the decay rate? Are supersharp
    resonances special in this respect?

\item Semiclassical approach. It seems that in order to reproduce resonance wave
    functions it is necessary to include more information than just periodic orbits
    (maybe diffraction). It is not clear how to proceed in this direction.

\item Statistics of eigenvectors. Virtually nothing is known in the case of
classical openings, even within random matrix theory.
\end{itemize}

\ack I gratefully acknowledge important feedback on a previous version of this work by
St\'ephane Nonnenmacher and Dmitry Savin, who kindly suggested some references and
provided some clarifications. Any imprecisions that may still linger are responsibility
of the author. During preparation of this manuscript I have enjoyed financial support
from FAPESP and CNPq.

\end{document}